\documentclass[twocolumn,showpacs]{revtex4}

\usepackage{graphicx}%
\usepackage{dcolumn}
\usepackage{amsmath}

\newcommand{\bb}{\begin{equation}}
\newcommand{\en}{\end{equation}}
\begin{document}

\title{Microrheology, stress fluctuations and active behavior of living cells}

\author{A.W.C. Lau$^{1}$}
\author{B.D. Hoffman$^{2}$}
\author{A. Davies$^{3}$}
\author{J.C. Crocker$^{2}$}
\author{T.C. Lubensky$^{1}$}
\affiliation{$^{1}$Department of Physics and Astronomy, University of
Pennsylvania, Philadelphia, PA 19104 \\
$^{2}$Department of Chemical and Biomolecular Engineering, University of Pennsylvania, Philadelphia, PA 19104 \\
$^{3}$Department of Applied Physics, California Institute of Technology,
Pasadena, CA 91125}
\date{\today}

\begin{abstract}
We report the first measurements of the intrinsic strain fluctuations of
living cells using a recently-developed tracer correlation technique along
with a theoretical framework for interpreting such data in heterogeneous
media with non-thermal driving.  The fluctuations' spatial and temporal
correlations indicate that the cytoskeleton can be treated as a
course-grained continuum with power-law rheology, driven by a spatially
random stress tensor field.  Combined with recent cell rheology results,
our data imply that intracellular stress fluctuations have a nearly
$1/\omega^2$ power spectrum, as expected for a continuum with a slowly
evolving internal prestress.
\end{abstract}
\maketitle

An accurate physical picture of the viscoelasticity and motion of the
cytoskeleton is crucial for a complete understanding of processes such
as intracellular transport \cite{motors},  cell crawling \cite{janmey},
and mechano-chemical transduction \cite{janmey}. Microrheology
\cite{microrhelogy}, based on the analysis of embedded tracer particle
motion, has recently emerged as an experimental probe of cytoskeleton
viscoelasticity and dynamics \cite{vivoexpt1,vivoexpt2,vivoexpt3,vitroexpts}.
The viscoelastic properties of eucaryotic cells arise from an intricate
network of protein filaments driven by specialized motor proteins and
directional polymerization, that convert the chemical energy of
adenosine triphosphate (ATP) to mechanical work and motion.  A cell is thus
a nonequilibrium soft material whose fluctuations are actively driven.
Unlike the thermal fluctuations in an equilibrium material, the amplitude and
spatial distribution of active fluctuations can be controlled via biochemical
signaling pathways; perhaps allowing the cell to locally adjust its' mechanical
properties to suit its' needs.  Indeed, microscopic force generators play a
central role in existing cell mechanics models such as the sol-gel \cite{solgel},
soft glassy rheology \cite{vivoexpt1} and tensegrity \cite{wang} hypotheses.

In this Letter, we extend a recently introduced method, termed
two-point microrheology \cite{john}, and show that it can be used to
characterize the activity of intracellular force generators by
directly measuring a cell's intrinsic, random stress fluctuations.
Our experimental data and theoretical framework show that a cell
can be modelled as a coarse-grained viscoelastic continuum driven
by a spatially random stress field having a $1/\omega^2$
power spectrum in our observable frequency range, $1 < \omega < 60
$ rad/s.

There are two distinct approaches to microrheology: the active approach
measures the displacements of tracer particles induced by external forces and
the passive approach measures fluctuations of particle positions in
the absence of driving forces.  The active approach provides a direct measure of
the complex shear modulus $\mu(\omega)$.  In equilibrium systems
the passive approach also measures $\mu(\omega)$ because of the
fluctuation-dissipation theorem (FDT) \cite{lubensky}.
Literature results in cells using single-particle versions of the
two approaches yield shear moduli differing by orders of magnitude
and exhibiting qualitatively different frequency dependencies \cite{vivoexpt1,vivoexpt3}.
These results are further limited by ambiguities
associated with tracer boundary conditions and medium heterogeneity and, more
importantly, the applicability of the FDT.   We show theoretically
that data from passive two-point microrheology and active response experiments can be combined to
measure the activity of molecular motors -- specifically, the power spectrum of
intracellular stress fluctuations -- despite the cell's heterogeneity.
We will first discuss current microrheology approaches and our measurements.

Microrheology relies on the Langevin equation \cite{lubensky}
for the velocity ${\bf v}$ of a tracer particle,\bb
m \partial_t {\bf v}(t) = -
\int_{-\infty}^{t}dt'\,\gamma(t-t')\,{\bf v}(t') +  {\bf f}_E(t)
+{\bf f}_R(t),
\label{GLE}
\en
where $m$ is the mass of the tracer particle,
${\bf f}_E(t)$ is the external force, and ${\bf f}_R(t)$
is the random force arising from the medium. Stokes law states
that $\gamma(\omega) = 6 \pi a\mu(\omega)/(-i \omega)$ \cite{micro2}
for spherical tracers with radius $a$ and no-slip
boundary conditions in an isotropic, homogeneous viscoelastic medium
with a complex shear modulus $\mu(\omega)$.  In the active scheme,
${\bf f}_R(t)$ can be set to zero, and the
displacement is ${\bf r}(\omega) = \chi(\omega)\,{\bf f}_E(\omega)$,
where $\chi(\omega) = \left \{ - i \omega \left [-i m\omega + \gamma(\omega) \right ] \right \}^{-1}$
is the response function.

Several measurements of $\chi(\omega)$ to extract
the elastic moduli of the cytoskeleton of living cells have been
performed \cite{vivoexpt1,vivoexpt2}.  In particular, Fabry {\em
et al}.\ \cite{vivoexpt1} have reported accurate measurements of the linear
response to an applied torque of a few-micron-diameter magnetic bead attached
to the outside of a cell and strongly coupled to the cytoskeletal network via
the cell's integrin receptors.  Remarkably, they found that over five decades
in frequency, from $10^{-2}$ to $10^3$ Hz, the shear modulus of several cell
types was of order $1$ kPa with a power law form
$|\mu(\omega)| \sim \omega^{\beta}$ with $0.10 < \beta <0.30$.

Passive `one-point' microrheology is based on the correlation
function $C_{r_i r_j} (t) =\langle r_i ( t ) r_j ( 0 )\rangle $.
In thermal equilibrium, this correlation function is related through
the FDT to the response function $\chi (\omega)$:
\begin{eqnarray}
C_{r_i r_j}(\omega)  = \delta_{ij} \frac{2 k_B T}{\omega}{\rm Im}\,\chi
( \omega)  \approx \delta_{ij} \frac{k_B T} {3 \pi
a}\frac{\mu''(\omega)}{\omega |\mu(\omega)|^2}.
\label{gsr}
\end{eqnarray}
The final form is valid in the window
$\omega_a < \omega < \omega_b$, with the lower frequency
$ \omega_a \sim 10^{-2}$ Hz set by the compressional mode of the network
and upper frequency $\omega_b \sim 10^5$ Hz by inertial effects \cite{alex}.
Experiments typically measure tracers' mean-squared-displacement (MSD),
$\langle \Delta {\bf r}^2 (\tau )\rangle = \langle \Delta r_i (t,\tau)
\Delta r_i(t,\tau) \rangle$, where $\Delta r_i(t,\tau) =
r_i(t+\tau) - r_i(t)$ and the brackets represent time and ensemble
averages. $\langle \Delta {\bf r}^2 (\tau )\rangle$ is related to the
correlation function by $\langle \Delta {\bf r}^2 (\tau )\rangle =
\int \frac{d \omega} {2\pi} (1 - e^{-i \omega \tau})\,C_{r_i r_i}( \omega )$.

Two-point passive microrheology experiments measure a two-particle displacement
tensor $D_{ij}(R,\tau) \equiv \langle \Delta {r_i^{(1)} (t,\tau)}
\Delta {r_j^{(2)} (t,\tau)} \rangle $, where the superscripts identify the
two different tracers, the brackets represent an ensemble average
over all tracer pairs and time, and $R$ is the separation between the
two tracers. In thermal systems, $D_{ij}(R,\tau)$ can be related via the FDT to the
shear moduli by two-particle equivalents of $C_{r_i r_j}(\omega)$,
but with the $1/a$ scaling replaced by $1/R$ scaling.  Importantly,
it has been shown \cite{alex,dan} that $D_{ij}(R,\tau)$ is independent of tracer size,
shape and boundary conditions (to leading order in $1/R$).  In cells, this allows the use of
endogenous particles, unlike existing methods that attach or inject synthetic
tracers which may perturb the cell.  In systems where Eq.\ (\ref{gsr})
is valid, $D_{ij}(R,\tau)$ and $\langle \Delta {\bf r}^2 (\tau )\rangle$
have the same functional form vs.\ $\tau$, differing only by a geometrical
constant.  This leads to the definition of a two-point MSD, which may be
thought of as an ideal Stokes particle advected by random fluctuations
of the medium:
$\langle \Delta {\bf r}^2 (\tau )\rangle_{2} \equiv (2R/a) D_{rr}(R,\tau)$,
where $D_{rr}$ is the tensor component of $D_{ij}$ parallel to $\hat{R}$.

We made passive microrheology measurements on two cultured cell lines
for which shear moduli $\mu(\omega)$ had been measured previously
\cite{vivoexpt1}.  All cells were cultured in Delbeco's Modified Eagle
Medium supplemented with $10$\% calf serum and $50$ mg/ml gentamicin in $5$\% CO$_2$.
F9 cells were cultured in gelatin-coated tissue-culture flasks.  Cells were passed
into glass-bottom cell-culture dishes with collagen-coated coverslips and
allowed to incubate for either 4-6 hours (F9) or overnight (J774A.1)
prior to experiments.  For tracers, we visualized endogenous refractive
particles in the cells using differential
interference contrast (DIC) microscopy and computed one- and two-particle
correlations from the same multi-particle video tracking trajectories
\cite{crocker}.  As our analysis assumes a three-dimensional continuum,
we excluded trajectories from the thin lamellar region of the cell and
the mechanically distinct nucleus, focussing instead on the midplane
of the 6-8$\mu$m thick cells.  We typically observed $\sim 100$ sub-micron
tracers (presumed by morphology to be lipid granules and mitochondria)
within our $1\,\mu\mbox{m}$ focal depth.  Thirty minutes of data yielded
$\sim 10^7$ tracer positions per cell, with respective time and space
resolution of $1/60$th sec and $20\,$nm.

\begin{figure}[tp]
{\par\centering
\resizebox*{3in}{2.1in}{\rotatebox{0}{\includegraphics{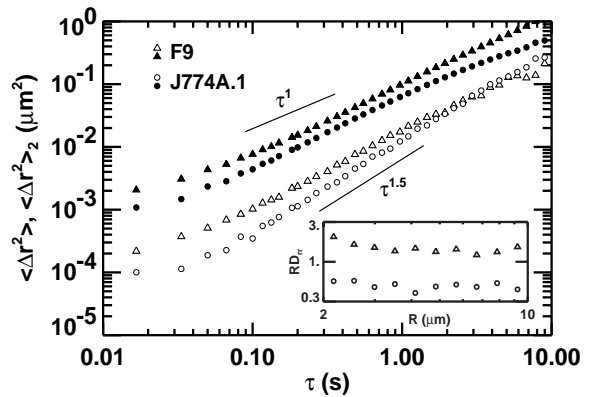}}}
\par}
\caption{ One (closed symbols) and two-particle (open symbols)
displacements vs.\ lag time $\tau$ for endogenous tracers in two types
of cultured cells, J774A.1 mouse macrophage and F9 mouse carcinoma.
Super-diffusive behavior, exponent greater than $1$, indicates the
effect of non-thermal fluctuations.  (inset) $RD_{rr}(R,\tau=0.067\,\mbox{sec})$
in units of $10^{-4} \mu{\rm m}^3$, is nearly constant in both cell types,
as expected for a coarse-grained isotropic continuum.}
\label{figure1}
\end{figure}

Typical MSD data for cells are presented in
Fig.\ \ref{figure1}.  Although visual examination of the images occasionally
shows ballistic tracer motion (trafficking), driven presumably by kinesin
and dynein motor proteins along microtubules, the ensemble averaged
$\langle \Delta {\bf r}^2 (\tau )\rangle $ is dominated by
random, apparently diffusive motion.  From the two-point data, we first
verified that $D_{rr} \sim 1/R$, [Fig.\ \ref{figure1} (inset)] in the
accessible range $2 < R < 8\,\mu$m,
comparable to results in other studies \cite{wang,bausch}.  In thermal
systems at least, this finding suggests that the medium may be treated
as a coarse-grained homogeneous continuum on the scale $R$.  Strikingly,
the two-point displacements of intracellular tracers exhibit a
super-diffusive behavior: $\langle \Delta {\bf r}^2 (\tau ) \rangle_{2}
\sim \tau^{\alpha}$ with exponent $1.30< \alpha < 1.60$.
If a living cell were an equilibrium viscoelastic medium with
$\mu(\omega) \sim \omega^{\beta}$ as reported in Ref.\ \cite{vivoexpt1},
Eq.\ (\ref{gsr}) would imply that $\langle \Delta {\bf r}^2(t) \rangle \sim t^\beta$.
Our data show that $\alpha > \beta$ and, thus, the FDT breaks down.
While the one-point MSD suggests that
the motion of single endogenous tracers is dominated by diffusive
motion relative to the network, which may be thermal or
non-thermal in origin, recent experiments have shown
$\langle \Delta {\bf r}^2(\tau) \rangle \sim \tau^{1.5}$ for large
(several $\mu$m) phagocytosed tracers \cite{vivoexpt3}.  Such one
particle measurements, however, can not distinguish between advection by a
driven continuum or trafficking relative to a stationary network.
In constrast, our two-point MSD results unambiguously indicate that
the cytoskeleton itself has large strain fluctuations driven by nonthermal forces.

\begin{figure}[tp]
{\par\centering
\resizebox*{3in}{2.1in}{\rotatebox{0}{\includegraphics{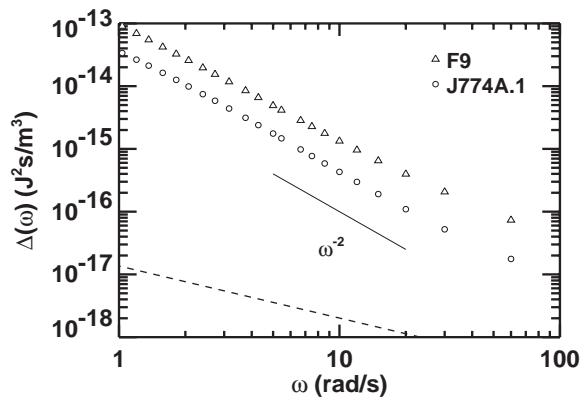}}}
\par}
\caption{The intracellular stress fluctuation spectrum $\Delta(\omega)$
as measured in two-point microrheology. The cell lines are J774A.1
macrophage and F9 carcinoma.  The dashed line represents the thermal
spectrum, $2 k_B T \mu'' ( \omega)/\omega$, where $\mu(\omega)$ is measured in
Ref.\ \cite{vivoexpt1}. }
\label{figure2}
\end{figure}

In living cells, active motors not only modify the viscoelastic response
of the cytoplasm, they also give rise to random, non-thermal stress
fluctuations that cause tracer particles to be subjected to random
nonthermal forces.  In such a system, Eq.\ (\ref{gsr}) must be modified.
We will outline a careful derivation below, but we can easily guess
it as follows: In thermal systems, the stress tensor has a random component
$s_{ij}({\bf x}, t)$ whose local fluctuations are described by\begin{eqnarray}
\langle s_{ij}({\bf x}, \omega) s_{kl}({\bf x}',\omega') \rangle
&=& 2 \pi \Delta(\omega) \delta(\omega +\omega') \delta^{3}({\bf x} - {\bf x}') \nonumber \\
&\times& \left [ \delta_{ik} \delta_{jl} + \delta_{il} \delta_{jk}
- { 2 \over 3}\,\delta_{ij}\delta_{kl} \right ], \label{stress}
\end{eqnarray}
where $\Delta(\omega) = 2 k_B T \mu'' ( \omega)/\omega$ as dictated by FDT
\cite{landau2}. In non-equilibrium systems, there are stress
fluctuations of the same form, but with $\Delta(\omega)$ not locked
to $\mu'' ( \omega)$ and we expect\bb
C_{r_i r_j}(\omega) \approx \delta_{ij} \frac{\Delta(\omega)}{6
\pi a |\mu(\omega)|^2},
\label{nonthermal}
\en
where $\Delta(\omega)$ is now interpreted as the power spectrum of the stress
fluctuations whose microscopic origin is the activity of the
motors.  Equation (\ref{nonthermal}) suggests that (i) tracers can
exhibit superdiffusive behavior
provided $\Delta(\omega)$ diverges sufficiently at small $\omega$,
and (ii) if an independent measure of $\mu(\omega)$ exists,  then $\Delta(\omega)$ can be sensibly
extracted from passive correlations data.

To avoid artifacts associated with
medium heterogeneity and non-Stokes' boundary conditions,
we used Eq.\ (\ref{drr}) below, which is the two-point equivalent of
Eq.\ (\ref{nonthermal}), along with rheological data reported
in Ref.\ \cite{vivoexpt1}, to convert our two-point data for both cell types
to $\Delta(\omega)$ as shown in Fig.\ \ref{figure2}.
We found a nearly $1/\omega^2$ spectra.
These typical results were replicated on $8$ cells of each type.
The variations of $\alpha$ and the power spectrum exponent among cells of
each type, and within different regions of a single cell, were comparable
within our measurement error.  A $1/\omega^2$
spectrum corresponds to a linear decay in time of a stress-stress correlation
function within our experimental time window, and would be a natural
consequence of slow evolution of intracellular stress \cite{time}.
Note also that $\Delta(\omega)$ is a few orders of magnitude greater
than the thermal spectrum (dashed line), affirming that these fluctuations
are driven by a nonthermal mechanism.

In the case that the shear modulus and the stress spectrum have power-law
forms, our result can be simply stated.  In living cells,
with `power-law' modulus $\mu(\omega) \sim \omega^{\beta}$ and a
$\Delta(\omega) \sim \omega^{-\gamma}$ spectrum, Eq.\ (\ref{nonthermal}) implies
that $\langle \Delta {\bf r}^2 (\tau) \rangle \sim  \tau^{\gamma + 2\beta -1}$.
If the $1/\omega^2$ spectrum we observe is universal, then one can measure
the rheology exponent $\beta$ via passive two-point measurements, using the
formula $\langle \Delta {\bf r}^2 (\tau)\rangle_{2} \sim
\tau^{1 + 2\beta}$.  The correspondence of our explicitly intracellular
results with those of Fabry {\em et al}.\ \cite{vivoexpt1} supports their
assertion that they are measuring intracellular rather than cortical
viscoelasticity.

Now, we turn to our theoretical task and address the
fundamental question: Given that living cells are highly heterogeneous
non-equilibrium systems in which the FDT does not apply,
do two-point passive microrheological experiments overcome ambiguities
associated with tracer boundary conditions and medium heterogeneity, and
in conjunction with response measurements, extract the power spectrum
of the continuum stress fluctuations?  We first postulate that a living cell at
large length scale effectively is an incompressible, viscoelastic
medium characterized by a displacement field ${\bf u}({\bf x},t)$,
whose equation of motion is given by \bb
\rho \,{\partial^2 u_i \over \partial t^2} =
\partial_j \sigma_{ij}({\bf x},t) + f_i({\bf x},t),
\label{medium}
\en
where $\rho$ is the coarse-grained mass density, $f_i({\bf
x},t)$ is the non-thermal random force density arising from the
motors, and $\sigma_{ij}({\bf x},t)$ is the stress tensor. Within
linear response theory, the stress $\sigma_{ij}({\bf x},t)$ is
related to the strain by $\sigma_{ij}({\bf x}, \omega) = 2
\mu(\omega) u_{ij}({\bf x}, \omega)$, where $u_{ij} = ( \partial_j
u_i + \partial_i u_j )/2$ is the strain tensor and $\mu(\omega)$
the frequency-dependent shear modulus.  Since there should be no net
external force in a stationary cell, the random force must take the form
$f_i({\bf x},t) = \partial_j s_{ij}({\bf x}, t)$, where $s_{ij}({\bf x},
t)$ is a random stress tensor.  We assume that its average is zero and variance
given by Eq.\ (\ref{stress}).  To relate to our microrheology
experiments, we relate one- and two-particle correlation functions
to stress fluctuations $\Delta(\omega)$ and response to
external forces $\mu(\omega)$ as follows.

We consider two tracer particles of radius $a$ placed in
this random medium and ask: What forces are exerted on each
particles?  For simplicity, we assume that the
heterogeneities near the vicinity of the tracers are well reflected
by a local effective shear modulus $\mu^{*}(\omega)$ which may be
different from $\mu(\omega)$ in the bulk.  Let particle 1 (2) at
${\bf x}$ (${\bf x}')$ undergo a displacement
$\varepsilon^{(1)}_i(\omega)$ ($\varepsilon^{(2)}_i(\omega)$).
First, we decompose ${\bf u}({\bf x},t )$ into
an average part and a fluctuating part: ${\bf u}({\bf x},t ) =
\bar{\bf u}({\bf x},t )+ \tilde{\bf u}({\bf x},t )$, and
solve Eq.\ (\ref{medium}) subject to boundary conditions: $ \bar{\bf u}_i(|{\bf x}| =a,
\omega) = \varepsilon^{(n)}_i(\omega)$ and $\tilde{\bf u}_i(|{\bf x}| =a , \omega) = 0$,
on the surface of the $n$th particle and $\bar{\bf u}_i(|{\bf x}|,\omega) $,
$\tilde{\bf u}_i(|{\bf x}|, \omega) \rightarrow 0$,
far away from the particles.   The total force exerted by the medium on each
particles has two components: an average,
$\bar{F}^{(n)}_i(\omega) = \int_{S_n} dS\,\hat{n}_j\,
\bar{\sigma}_{ij}({\bf x},\omega)$, and a random
parts, $\tilde{F}^{(n)}_i(\omega) = \int_{S_n}
dS\,\hat{n}_j\,\tilde{\sigma}_{ij}({\bf x},\omega)$, where
$\bar{\sigma}_{ij} = 2 \mu(\omega) \bar{u}_{ij}$ and
$\tilde{\sigma}_{ij} = 2 \mu(\omega) \tilde{u}_{ij}$ are,
respectively, the average and fluctuating stress, and $\hat{n}_j$
is the unit surface normal pointing towards the center of each
particle.  It is straightforward to compute $\bar{F}^{(n)}_i(\omega) =
\sum_{m} \chi^{-1\,(m,n)}_{ij}(\omega) \varepsilon^{(m)}_i(\omega)$,
where $\chi^{(n,m)}_{ij}(\omega)$ is the two-particle response
matrix given by $\chi^{(1,1)}_{ij}(\omega) = {
\delta_{ij}\over  6 \pi a \mu^{*}(\omega)}$ and
$\chi^{(1,2)}_{ij}(\omega) = { \hat{R}_i \hat{R}_j \over  4 \pi R
\mu(\omega)} + { \delta_{ij} - \hat{R}_i \hat{R}_j \over 8 \pi R
\mu(\omega)}$, to the lowest order in $1/R$, where $R \equiv |{\bf
x} - {\bf x}'|$, the distance between the two particles and
$\hat{\bf R} \equiv {\bf R}/R$; the noise correlators are given by
$\langle \tilde{F}^{(n)}_i(\omega)\,\tilde{F}^{(m)}_j(-\omega) \rangle =
{\Delta(\omega) \,\chi^{-1\,(n,m)}_{ij}(-\omega) / \mu(-\omega)}$ \cite{fluctuating}.
These results imply that $\langle \varepsilon^{(1)}_i(\omega) \varepsilon^{(1)}_j(-\omega) \rangle$
depends in a complicated way on $\mu^{*}(\omega)$ and $\mu(\omega)$.
In contrast, the cross-correlation
function as measured by two-point microrheology is\bb
D_{rr}(R, \omega) \equiv \langle \varepsilon^{(1)}_i(\omega) \varepsilon^{(2)}_i(-\omega) \rangle =
{ \Delta(\omega) \over 6 \pi R \,|\mu(\omega)|^2 },
\label{drr}
\en
to the lowest order in $1/R$, which depends only on $\mu(\omega)$ and
$\Delta ( \omega )$ in the bulk, and is independent of the
tracers' size, shape or boundary conditions.  Thus, apart from a geometrical
factor, Eqs.\ (\ref{nonthermal}) and (\ref{drr}) are equivalent, and this
justifies our interpretation of our experiments.  Furthermore, the $1/R$
behavior shown in Fig.\ \ref{figure1} (inset) not only implies that the
living cell can be treated as a continuum, but also requires that the random
force $f_i({\bf x},t)$ arises from a stress
tensor of the form given in Eq.\ (\ref{stress}).

Lastly, we propose a microscopic picture of motor
activities and derive a fluctuating stress tensor.
Since motors are small but finite objects, their
activities disturb the ambient cytoplasm in the form of a point
dipole \cite{rama}.  Since there is no net external force inside a stationary
cell, by Newton's 3rd law, the force exerted on the cytoplasm by a
motor must be equal and opposite to that of the cytoplasm on that
motor.  Thus, the part of the stress tensor arising from deviations
$\delta c_{a}({\bf x}, t)$ of the coarse-grained activity density of
motors from its average $c_a$ is $ s_{ij}({\bf
x}, t) = \Gamma \left ( \hat{n}_i \hat{n}_j - { 1\over 3}
\delta_{ij} \right ) \delta c_{a}({\bf x}, t),$ where ${\bf \hat{n}}$ is the
direction of the point dipole and $\Gamma$ is the energy scale.
Since there is no preferred direction, we can average over all angles to obtain
\begin{eqnarray} \langle s_{ij}({\bf x}, \omega) s_{kl}({\bf x}',
-\omega) \rangle &=& { \Gamma^2 \over 15}  \left [ \delta_{ik}
\delta_{jl} + \delta_{il} \delta_{jk} -
{ 2 \over 3}\, \delta_{ij}\delta_{kl} \right ] \nonumber \\
&\times& \langle \delta c_{a}({\bf x}, \omega) \delta c_{a}({\bf x}', -\omega) \rangle,
\end{eqnarray}
which has exactly the same form of Eq.\ (\ref{stress}) if
the fluctuations of motor activity is delta correlated in space.
Furthermore, we can estimate the
power associated with the stress fluctuations by computing
the power dissipated per tracer:
$\langle P \,\rangle = \int { d\omega \over 2 \pi}\,\omega\,\mu''(\omega)\,
\Delta(\omega)/\left | \mu(\omega) \right |^2 \sim 10^{-16}\,\mbox{W}$.
Estimating $ \sim 10^3$ tracers per cell, the power to drive one cells'
fluctuations is of the order of $10^{-13}\,\mbox{W}$, which,
as $1\%$ of typical cell metabolism, is not unreasonable.

In conclusion, the interpretability of two-point microrheology in complex
media, including cells, serves to clarify an otherwise confusing set of
biophysical observations, supporting efforts to model cells as
three-dimensional continua rather than cortical shells and suggesting
that the cytoskeleton is a highly dynamic, actively stressed network.
Future work to extend the temporal range and statistical power of
such measurements should enable mapping and non-trivial spectroscopy of
intracellular rheology and stress.  Finally, such techniques may prove fruitful in
other systems with non-thermal fluctuations, such as
granular and jamming media.

We thank Mark Goulian, Paul Janmey and Phil Nelson for valuable discussions.
This work is supported by the NSF through the MRSEC Grant DMR 00-79909.
JCC acknowledges support from the David and Lucile Packard Foundation.


\begin{thebibliography}{0}%


\bibitem{motors}
J. Howard, {\em Mechanics of Motor Proteins and the Cytoskeleton},
(Sinauer, New York, 2000).

\bibitem{janmey}
P.A. Janmey, Physiol.\ Rev.\ {\bf 78}, 763 (1998).

\bibitem{microrhelogy}
Y. Tseng {\em et al.}, Current Opinion in Coll. \& Interf. Sci. {\bf 7}, 210 (2002).

\bibitem{vivoexpt1}
B. Fabry {\em et al.}, Phys. Rev. Lett. {\bf 87}, 148102 (2001).

\bibitem{vivoexpt2}
P.A. Valberg and H.A. Feldmen, Biophys. J. {\bf 52} 551 (1987); A.R. Bausch {\em et al.},
Biophys. J. {\bf 76} 573 (1999); S. Yamada {\em et al.}, Biophys. J. {\bf 76} 1736 (2000).

\bibitem{vivoexpt3}
A. Caspi {\em et al.}, Phys. Rev. Lett. {\bf 85}, 5655 (2000); Phys. Rev. E {\bf 66}, 011916 (2002).

\bibitem{vitroexpts}
L. Le Goff {\em et al.} Phys. Rev. Lett. {\bf 88}, 018101 (2002).

\bibitem{solgel}
P.A. Janmey {\em et al.}, Nature {\bf 345}, 89 (1990).

\bibitem{wang}
N. Wang {\em et al.}, Proc. Natl. Acad. Sci. USA {\bf 98}, 7765 (2001).

\bibitem{john}
J.C. Crocker {\em et al.}, Phys. Rev. Lett. {\bf 85}, 888 (2000).

\bibitem{lubensky}
P.M. Chaikin and T.C. Lubensky, {\em Principles of Condensed Matter Physics}
(Cambridge University Press, New York, 1995).

\bibitem{micro2}
T.G. Mason, D.A. Weitz, Phys. Rev. Lett. {\bf 74} 1250 (1995);
F.C. MacKintosh, C.F. Schmidt, Current Opinion in Coll. \& Interf. Sci. {\bf 4} 300 (1999).

\bibitem{alex}
Alex J. Levine and T.C. Lubensky, Phys. Rev. Lett. {\bf 85}, 1774
(2000); Phys. Rev. E {\bf 63}, 041510 (2001); Phys. Rev. E {\bf 65}, 011501 (2001).

\bibitem{dan}
D.T. Chen {\it et al.}, Phys. Rev. Lett. {\bf 90}, 108301 (2003);
L. Starrs and P. Bartlett, Faraday Discuss. {\bf 123}, 323 (2003).

\bibitem{crocker}
J.C. Crocker and D.G. Grier,  J. Colloid Interface Sci. {\bf 179},  298 (1996).

\bibitem{bausch}
A.R. Bausch {\it et al.}, Biophys. J. {\bf 75}, 2038 (1998).

\bibitem{landau2}
L.D. Landau and E.M. Lifshitz, {\em Statistical Physics Pt. 2}, 2nd ed. (Pergamon Press, Oxford, 1980).

\bibitem{time}
The stress generation/relaxation may rely on a number
of modes with diverse time scales, $\tau_i$.  In the simplest
case, a stress autocorrelation would then be multi-exponential,
consistent with our result if all $\tau_i$ lie well outside of our
measurable range.

\bibitem{fluctuating}
Details will be presented elsewhere.

\bibitem{rama}
R.A.\ Simha and S.\ Ramaswamy, Phys. Rev. Lett. {\bf 89}, 058101 (2002).


\end{thebibliography}
\end{document}